# Prediction of new Group IV-V-VI monolayer semiconductors based on first principle calculation

Qingxing Xie[a], Junhui Yuan[a,b], Niannian Yu[a], Lisheng Wang[a], Jiafu Wang[a,c,*]

[a] Department of Physics and Institute of Applied Condensed Matter Physics, School of Science, Wuhan University of Technology, Wuhan, Hubei 430070, China

[b] School of Optical and Electronic Information, Huazhong University of Science and Technology, Wuhan, Hubei 430074, China

[c] Hubei Engineering Research Center of RF-Microwave Technology and Application, Wuhan University of Technology, Wuhan, Hubei 430070, China

* Corresponding author, email: jasper@whut.edu.cn (Jiafu Wang).

**Abstract:** Two-dimension (2D) semiconductor materials have attracted much attention and research interest for their novel properties suitable for electronic and optoelectronic applications. In this paper, we have proposed an idea in new 2D materials design by using adjacent group elements to substitute half of the atoms in the primitive configurations to form isoelectronic compounds. We have successfully taken this idea on group V monolayers and have obtained many unexplored Group IV-V-VI monolayer compounds: $P_2SiS$, $As_2SiS$, $As_2GeSe$, $Sb_2GeSe$, $Sb_2SnTe$, and $Bi_2SnTe$. Relative formation energy calculations, phonon spectrum calculations, as well as finite-temperature molecular dynamics simulations confirm their stability and DFT calculations indicate that they are all semiconductors. This idea broadens the scope of group V semiconductors and we believe it can be extended to other type of 2D materials to obtain new semiconductors with better properties for optoelectronic and electronic applications.

**Keywords:** new group IV-V-VI monolayers, DFT calculation, semiconductor

## Introduction

Since graphene was firstly exfoliated in 2004[1], two-dimensional (2D) materials have attracted tremendous research interest because of its unique properties and great potential for device applications[2-4]. This achievement has shown that it is possible to exfoliate stable, single-atom or single-polyhedron thick 2D materials from van der Waals solids with unique and fascinating physical properties. However, many 2D materials have shortcomings that limit their applications. For example, in graphene, despite its incredibly high carrier mobility, the lack of bandgap (spin-orbit coupling merely opens a gap of 0.003 meV at the Dirac cone) strictly limit its applications[5].

MoS$_2$ has a direct bandgap of 1.8eV and shows great potential in nanoelectronics, optoelectronics, and flexible device[6-8], but the carrier mobility of MoS$_2$ is not high enough due to its less dispersive band edges with relatively localized transition metal d character[9-10]. Therefore, much effort has been devoted to improve the performances of these existing 2D materials through different approaches, such as doping, functionalization, defects, and hybrid structures. Meanwhile, researchers are expecting to explore new 2D materials with desired properties.

In analogy to graphene, group IV elemental monolayers such as silicene and germanene have been proposed these years[11-14]. Cahangirov *et al.* reported that monolayer silicon and germanium can have stable, slightly corrugated, honeycomb structures which are more stable than their corresponding planar-layer counterparts[15]. Both low-buckled silicene and germanene are semimetals with Dirac cones at K points, and the Fermi velocities were estimated to be $10^6$ m/s, very close to graphene[16]. Recent years, group V elemental monolayers have stepped into researchers' horizon since monolayer black phosphorus (Phosphorene) has been fabricated successfully for the first time[17-18]. Phosphorene is a highly anisotropic 2D elemental material with promising semiconductor properties for flexible electronics[19]. Unlike transition metal dichalcogenides (TMDs), the carrier mobility in phosphorene (about 1000 cm$^2$V$^{-1}$s$^{-1}$)[20] has been found to be significantly higher than TMDs and the drain current modulation is up to $10^5$ [21], which makes phosphorene a potential candidate for future nanoelectronic applications. However, poor stability limits its applications and many aspects of the material still remain to be elucidated[22]. Recently, blue phosphorus, one of allotropes in phosphorus, which shares similar structure with silicene, has been reported to be a Dirac material after being hydrogenated or fluorinated[23]. This interesting result led researchers to consider other group V monolayer materials with buckled honeycomb lattice, and arsenene and antimonene have been predicted by first principle calculations[24-25].

A useful way to get new materials with similar configurations is to find isoelectronic compounds. This approach in new materials design is defined as replacing one type of atoms by neighboring group elements in the periodic table while keeping the total number of valence electrons unchanged. As the matter of fact, group III−V monolayer compounds with honeycomb structures, say h-BN, can be treated as an isoelectronic compound of graphene by substituting the two C atoms in the cell by one B atom and one N atom[26]. As monolayer black phosphorus has been discovered and many other group V monolayers proposed based on DFT calculation, their corresponding isoelectronic group IV-VI monolayer compounds such as SnS, SnSe, and GeTe have attracted plenty of attention[27-28]. For example, Zhu *et al*. have predicted a new 2D material SiS with good kinetic stability, which is obtained by replacing the two P atoms in the cell of monolayer blue phosphorus by one Si atom and one S atom [29].

In 2016, a new compound of group III-IV-V monolayer $Si_2BN$ was predicted[30]. By analyzing the crystal structure of this monolayer compound, we realize that it can be treated as an isoelectronic compound of silcence by replacing half of the Si atoms by neighboring elements of the B and N atoms. This triggered us to a new idea in the new material design. Different from using adjacent group elements to completely substitute all the atoms in the primitive cell, the idea is to half replace the atoms in the primitive system with neighboring elements on the periodic table to form corresponding isoelectronic compounds. In this paper, we will examine whether this idea can work for group V monolayers such as phosphorene (blue phosphorus structure), arsenene, and antimonene (their structures can be seen in Ref.[24]). Based on phosphorene, for example, a new isoelectric compound of monolayer $P_2SiS$ is designed where Si and S atoms are chose to half replace the P atoms. The structure of this isoelectric compound consists of P-P-Si-S alignment in a single layer in which each P atom has one P, one Si, and one S nearest neighbors, while each Si (S) has two P atoms and one S (Si) atom as nearest neighbors. For arsenene, not only Ge and Se atoms are chose to design the new monolayer As2GeSe, but also the former periodic elements Si and S atoms are taken into consideration ($As_2SiS$). For antimonene, which is similar to arsenene, monolayer $Sb_2GeSe$ and $Sb_2SnTe$ are designed. For monolayer bismuth, $Bi_2PbPo$ are not included in our discussion because of the radioactivity of polonium, and only $Bi_2SnTe$ is investigated.

It appears that partial substitution in isoelectronic compounds may result in 2D semiconducting IV-V-VI compounds that are stable and with excellent properties for various applications. In this work, we have systematically studied the structural stability and electronic properties of the designed layered structures discussed above ($P_2SiS$, $As_2SiS$, $As_2GeSe$, $Sb_2GeSe$, $Sb_2SnTe$ and $Bi_2SnTe$). Calculation results show that they are all semiconductors and are dynamically and thermodynamically stable.

**Methods**

Our first principle calculations are performed by Atomistix ToolKit version 2016.1

package[31-32] which is based on LCAO-based Density Functional Theory (DFT). The exchange-correlation functional is treated using the generalized gradient approximation (GGA-PBE) functional. Electron static potential is obtained by solving Poisson equation using FFT method. Double Zeta Polarized basis set is used for all atoms. The real-space mesh cutoff is 75 Ry, and the electronic temperature used in the Fermi distribution is set to be 300 K. A vacuum layer of more than 20 Å is used to simulate the isolated two-dimensional structures. All structures are fully optimized based on the Broyden-Fletcher-Goldfarb-Shanno (BFGS) method without any symmetry constraints. The structures are relaxed until the atomic forces are less than 0.005eV/Å and total energies are converged to $10^{-6}$ eV. A 11×11×1 grid sampling with Gamma-centered Monkhorst-Pack grid method are adopted for structure optimization, while a 21×21×1 grid sampling with the same grid method are used for the property calculation. The electronic iterations convergence criterion was set to be $10^{-7}$ eV.

Since the band gaps may be underestimated by DFT, we also carried out calculations using the screened hybrid functional, HSE06 method. In addition to relative formation energy and phonon spectra calculations, the stability of the optimized structures are verified at room temperature (300K) by performing ab initio finite-temperature molecular dynamics calculations. The Nose Thermostat is used and Newton's equation of motion is integrated through the Verlet algorithm with time steps of 2 fs.

**Results and discussion**

Through full optimization, we find that the structure models of the six 2D materials described above share the same features as shown in Figure 1. They all remain honeycomb lattice which is similar to graphene and $Si_2BN$[30]. However, different from the graphene's structure with no out-of-plane buckling, all of our designed 2D materials adopt buckled structures to increase their stability (See in Figure 1). The unit cell is a hexagonal. Each unit has 8 atoms: 4 group-V, 2 group-IV, and 2 group-VI elements. Optimized structural parameters of $P_2SiS$, $As_2SiS$, $As_2GeSe$, $Sb_2GeSe$, $Sb_2SnTe$, and $Bi_2SnTe$ are listed in Table 1, where *a* is the lattice constant, *h* is the buckling constant defined as the distance of two adjacent group V elements along c direction (see Figure 1), *d* is the bond length between two adjacent group V elements (P, As, Sb，or Bi), and $l_{V-IV}$, $l_{V-VI}$, and $l_{IV-VI}$ are bond lengths between V-IV elements, V-VI elements, and IV-VI elements, respectively. One can see that all the lattice constant, the buckling constant, and the bond lengths increase monotonously with respect to increasing atomic number expect for As$_2$GeSe, whose buckling constant and the bond length between two adjacent group V elements are slightly smaller than that of As$_2$SiS.

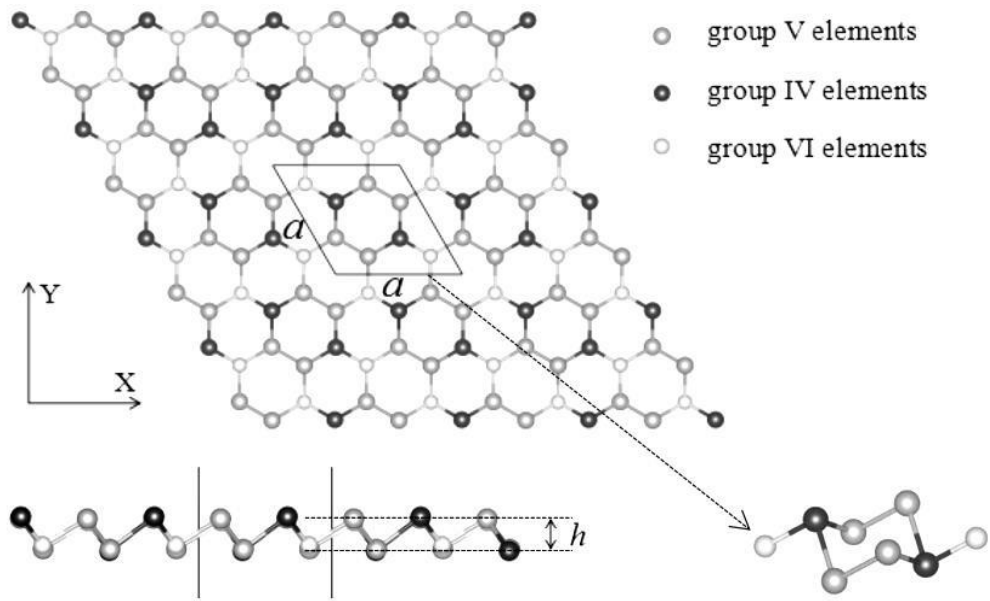

Figure 1. Lattice configurations (top views and side views) of P$_2$SiS, As$_2$SiS, As$_2$GeSe, Sb$_2$GeSe, Sb$_2$SnTe, and Bi$_2$SnTe monolayers. Their artuctures are similar to each other, so only one model is shown here.

Table 1. Structural parameters (unit in Å) and relative formation energy (unit in meV) of P$_2$SiS, As$_2$SiS, As$_2$GeSe, Sb$_2$GeSe, Sb$_2$SnTe and Bi$_2$SnTe.

| Structure | a | h | d | $l_{V-IV}$ | $l_{V-VI}$ | $l_{IV-VI}$ | ΔE |
|---|---|---|---|---|---|---|---|
| P$_2$SiS | 6.6201 | 1.1667 | 2.2441 | 2.3890 | 2.2413 | 2.3810 | 92 |
| As$_2$SiS | 6.9540 | 1.4142 | 2.5002 | 2.5030 | 2.3812 | 2.3551 | 66 |
| As$_2$GeSe | 7.2817 | 1.3617 | 2.4771 | 2.5931 | 2.5554 | 2.6395 | 58 |
| Sb$_2$GeSe | 7.8169 | 1.5856 | 2.8613 | 2.7812 | 2.7415 | 2.6474 | 49 |
| Sb$_2$SnTe | 8.2800 | 1.6176 | 2.8814 | 2.9431 | 2.9124 | 2.9694 | 38 |
| Bi$_2$SnTe | 8.4936 | 1.6855 | 3.0266 | 3.0167 | 2.9847 | 2.9393 | 36 |

Stability and experimental feasibility are important for new 2D materials designed. Geometry optimization calculations show that all these 2D materials present energetically stable configurations. Firstly, we calculated the relative formation energy (ΔE) of these 2D materials (listed in Table 1). ΔE is defined as

$$\Delta E = \frac{E_{2D}}{N_{2D}} - \frac{E_{3D}}{N_{3D}}$$

where $E_{2D}$ and $E_{3D}$ are the total energies of the 2D and 3D structures, respectively, and $N_{2D}$ and $N_{3D}$ are the numbers of atoms in the 2D and 3D unit cells[33]. We find that their relative formation energies are all less than 100meV/atom, conforming to the criterion proposed in Ref. [34]. Furthermore, we also performed the ab initio molecular dynamics (AIMD) simulations on these 2D materials to check their thermodynamic stabilities. As shown in Figure 2, after 1500 time steps (3ps), both the $P_2SiS$ and $As_2SiS$ are just slightly distorted; so are the other four monolayers (see Supplementary Material, Figure S1).

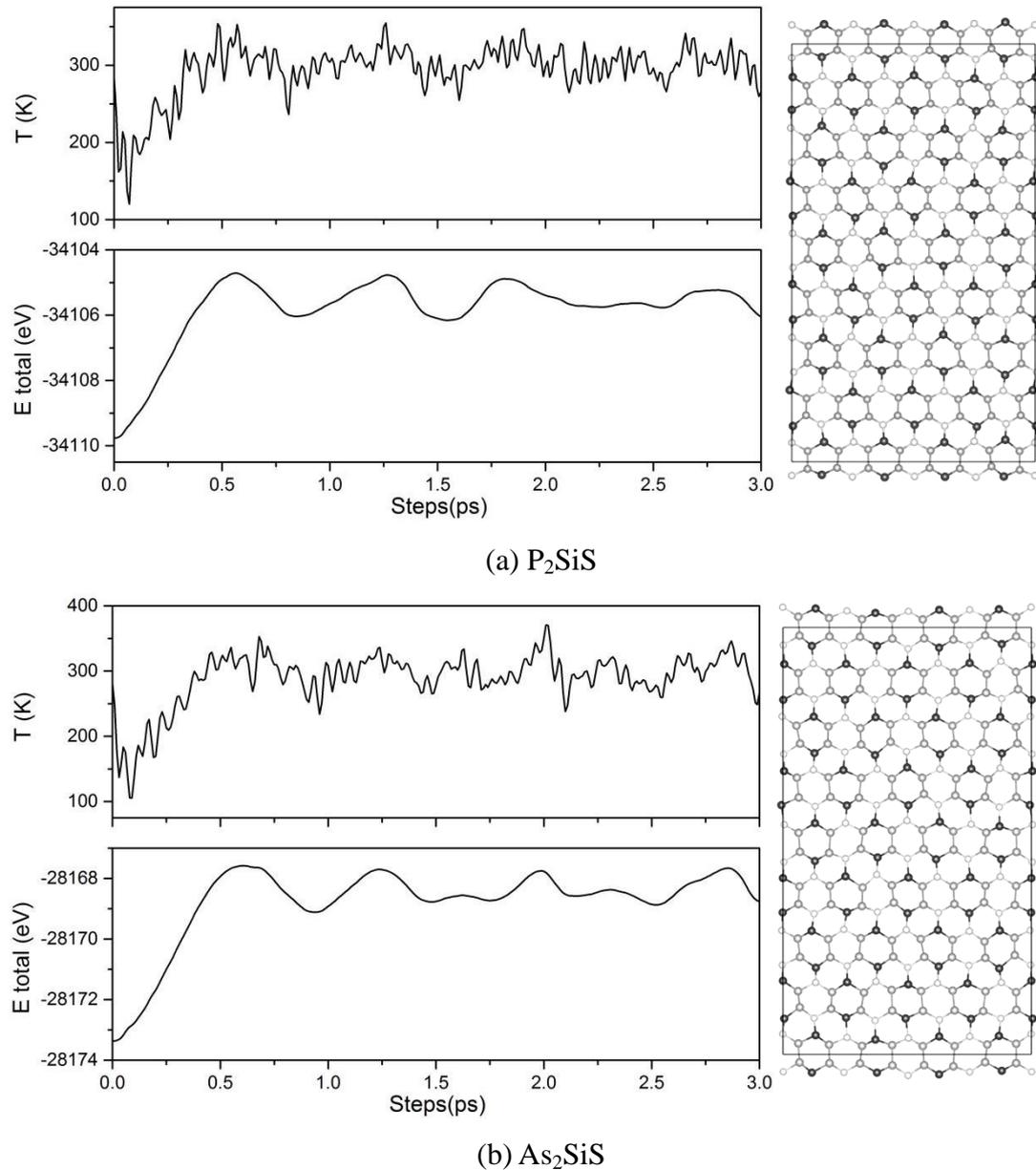

(a) $P_2SiS$

(b) $As_2SiS$

Figure 2. The fluctuations of temperature and the free energy (left) and the final structures (right) of the AIMD simulations for the $P_2SiS$ (a) and $As_2SiS$ (b) monolayers. The calculation results for the rest of four monolayers can be seen in Supplementary Material.

To further verify their stability, we have checked their phonon spectra. Phonon band structure calculation gives a reliable criterion to judge the stability -- the structure is stable only if all phonon frequencies are positive. We calculated phonon spectra through the finite displacement method. As shown in Figure 3, for $P_2SiS$ and $As_2SiS$ as examples, no imaginary frequency occurs on their phononic dispersion curves, showing that their structures are dynamically stable. The same phenomenon is found on the rest of four monolayers (see Supplementary Material, Figure S2), indicating they are all stable. Through these works, not only do we justify the dynamic and thermodynamic stabilities of these structures, but also the idea of using corresponding isoelectronic compounds with half replacing has been successfully extended to group V monolayers, demonstrating a useful way to broaden 2D materials design.

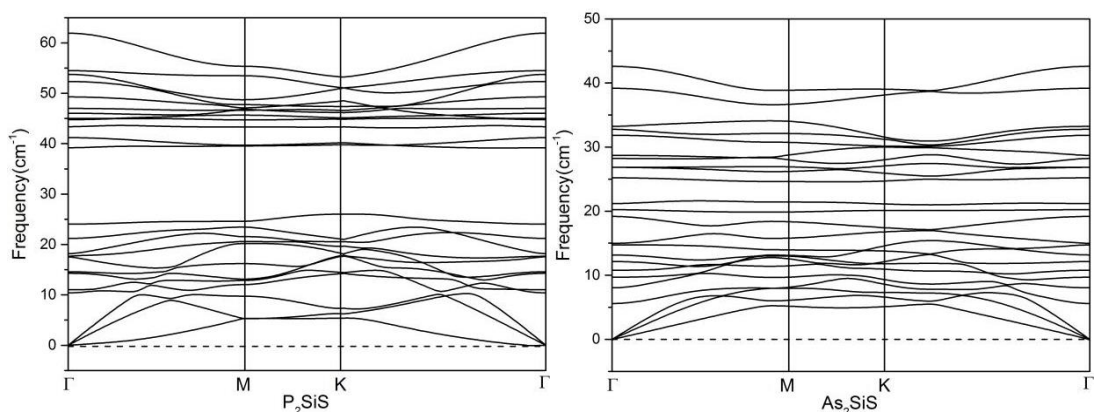

Figure 3. Phonon band structure of $P_2SiS$ and $As_2SiS$. Dotted line mark zero point of vibration frequency. The phonon spectrum for the rest of four monolayers can be seen in Supplementary Material.

Besides their stability, we have also investigated the electronic structures of the six designed monolayers. Figure 4 shows the band structures and density of states (DOS) for $P_2SiS$ and $As_2GeSe$ obtained by DFT calculation with GGA-PBE pseudopotentials. The electronic structures for the other four monolayers can be found in Supplementary Material (Figure S3 and S4). As can be seen in the figures, all the monolayers are direct bandgap semiconductors expect for $P_2SiS$ and $As_2SiS$. For the four direct bandgap monolayers ($As_2GeSe$, $Sb_2GeSe$, $Sb_2SnTe$, and $Bi_2SnTe$), all of their valence band maxima and conduction band minima are located at the Γ points. Comparing their bandgaps with each other, we find that the bandgaps of the four direct bandgap semiconductors decrease monotonously with respect to increasing atomic numbers ($P_2SiS$ and $As_2SiS$ are indirect bandgap semiconductors and are not comparable to them). Corrections to DFT band energies using the HSE06 method confirmed this situation. The HSE06 results indicate that $As_2GeSe$, $Sb_2GeSe$, $Sb_2SnTe$, and $Bi_2SnTe$ are direct bandgap semiconductors with bandgap of 1.464, 1.306, 1.199, 1.161eV respectively, while $P_2SiS$ and $As_2SiS$ are indirect bandgap semiconductors with bandgaps of 1.546 and 1.567eV. Moreover, we find that the tops of valence band are flat for all of the six monolayers, resulting in heavy hole masses and large DOS in the corresponding regions (see Figure 4, as well as Figure S3 and S4 in Supplementary Material). Their partial densities of states indicate that the valence band maxima and conduction band minima are mainly attributed to p state electrons of all the atoms, while the s electronic states contribute mainly to deep energy levels of the bands.

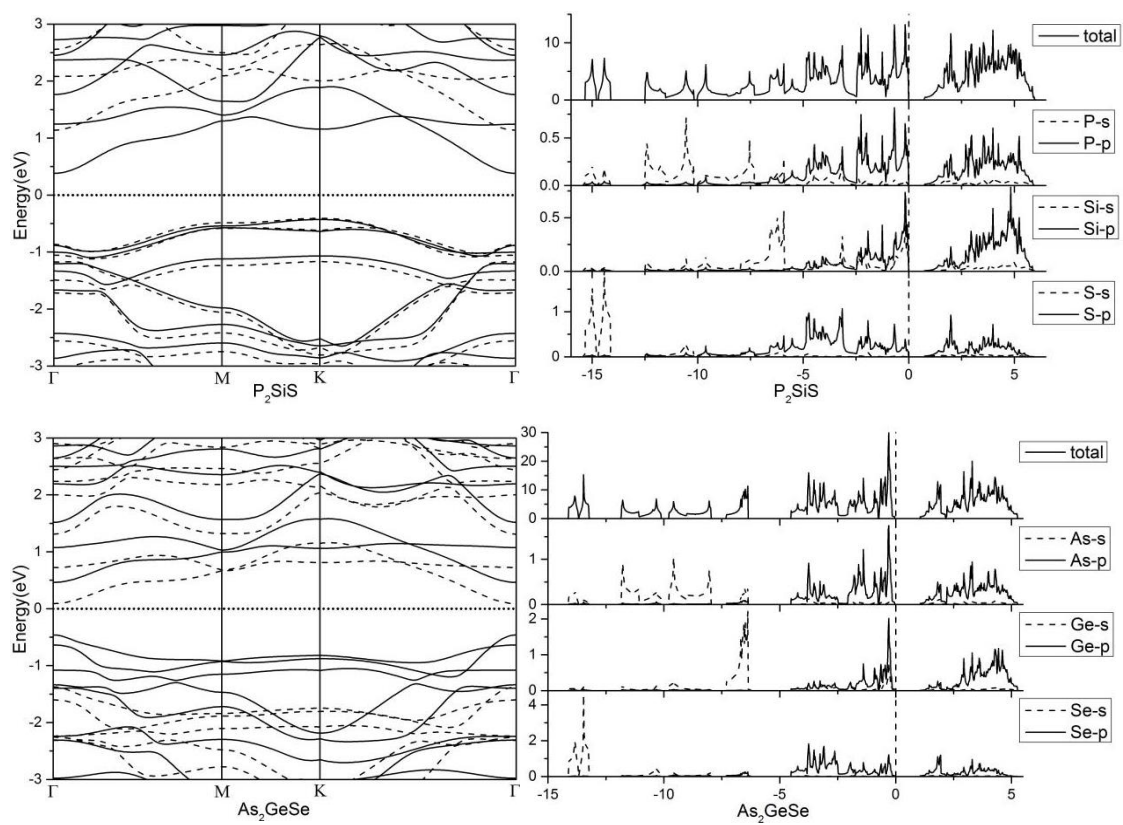

Figure 4. Band structures (left) and [partial] density of states (right) for $P_2SiS$, and $As_2GeSe$. The horizontal dotted lines (in band structure diagrams) and vertical dotted lines mark (in DOS diagram) represent Fermi levels. The electronic energy bands are calculated by both GGA-PBE (solid lines) and HSE06 (dashed lines) methods.

**Conclusion**

In this paper, we have proposed six new designed group IV-V-VI monolayer compounds. Based on first principle calculations, we have demonstrated their stabilities and then have explored the electronic structures for $P_2SiS$, $As_2SiS$, $As_2GeSe$,

$Sb_2GeSe$, $Sb_2SnTe$, and $Bi_2SnTe$. That these monolayers are stable is concluded not only based on the phonon band structure calculations, which demonstrate their dynamic stabilities, but also based on verifying their thermodynamic stabilities achieved by relative formation energy calculation and AIMD simulations. Their electronic band structures indicate that the monolayers $P_2SiS$ and $As_2SiS$ are indirect bandgap semiconductors, while the monolayers $As_2GeSe$, $Sb_2GeSe$, $Sb_2SnTe$, and $Bi_2SnTe$ are direct bandgap semiconductors, with gap values decreasing monotonously when increasing atomic numbers. More importantly, we have proposed an idea in new 2D materials design. The primitive cell can be transformed by substituting half of its primitive atoms by adjacent group elements, instead of completely replacing all primitive atoms in the cell, to obtain isoelectronic compounds. With this idea we are able to broaden the scope of group-V based semiconductors. We believe that this idea can be extended to other type of 2D materials so as to find new monolayers and/or multilayers with better properties for various optoelectronic and electronic applications.

**Acknowledgments**

This work was supported by the National Natural Science Foundation of China (Grant No. 11504281) and the Fundamental Research Funds for the Central Universities (Grant No. 2016-zy-067 and Grant No. 2015-Ia-008).